\documentclass[aps,prl,twocolumn,superscriptaddress]{revtex4-1}

\usepackage{mathrsfs}
\usepackage{amsmath}
\usepackage{amssymb}
\usepackage{amstext}
\usepackage{graphicx, epsf}
\usepackage{enumerate}
\usepackage{bbm}
\usepackage{color}
\usepackage[english]{babel}

\begin{document}

\title{Emergent irreversibility and entanglement spectrum statistics}

\author{Claudio Chamon}

\affiliation{Department of Physics, Boston University, Boston,
  Massachusetts 02215, USA}

\author{Alioscia Hamma}

\affiliation{Center for Quantum Information, Institute for
  Interdisciplinary Information Sciences, Tsinghua University, Beijing
  100084, P.R. China}

\author{Eduardo R. Mucciolo}

\affiliation{Department of Physics, University of Central Florida,
  Orlando, Florida 32816, USA}

\date{\today}

\begin{abstract}
We study the problem of irreversibility when the dynamical evolution
of a many-body system is described by a stochastic quantum
circuit. Such evolution is more general than a Hamiltonian one, and
since energy levels are not well defined, the well-established
connection between the statistical fluctuations of the energy spectrum
and irreversibility cannot be made. We show that the entanglement
spectrum provides a more general connection. Irreversibility is marked
by a failure of a disentangling algorithm and is preceded by the
appearance of Wigner-Dyson statistical fluctuations in the
entanglement spectrum. This analysis can be done at the wave-function
level and offers an alternative route to study quantum chaos and
quantum integrability.
\end{abstract}

\maketitle


In closed quantum systems, evolution is unitary and both
irreversibility and nonintegrability are elusive notions. Because of
unitarity, evolution is always stable under errors in initial
conditions. Thus, in quantum mechanics irreversibility is defined by
the vanishing of the probability (known as fidelity) of returning to
an initial state under arbitrarily small imperfections in the
Hamiltonian during the reversed time evolution
\cite{Peres84}. Nonintegrability is associated to a Wigner-Dyson
distribution of the energy-level spacings that shows level repulsion
\cite{gutzwiller} and nonintegrable Hamiltonians in this context are
irreversible. Integrable Hamiltonians, instead, tend to show
clustering of energy levels but can be either reversible or
irreversible \cite{znidaric,benenti}. When the time evolution is not
governed by a Hamiltonian, or when the Hamiltonian is time dependent,
energy levels are not well defined and these associations cease to be
meaningful. How can one relate nonintegrability and irreversibility in
these more general cases of quantum evolution?

In this Letter we show that one can answer this question by looking
at the wave function alone. This route allows one to study generic
quantum evolutions even when energy is not well defined. We show that
by studying the level statistics of the entanglement spectrum one can
determine whether the evolution is irreversible or not through a
protocol that we call entanglement cooling. It turns out that the
onset of irreversibility is marked by the presence of Wigner-Dyson
statistics in the entanglement spectrum.

\begin{figure}
\centering
\scalebox{.22}{\includegraphics{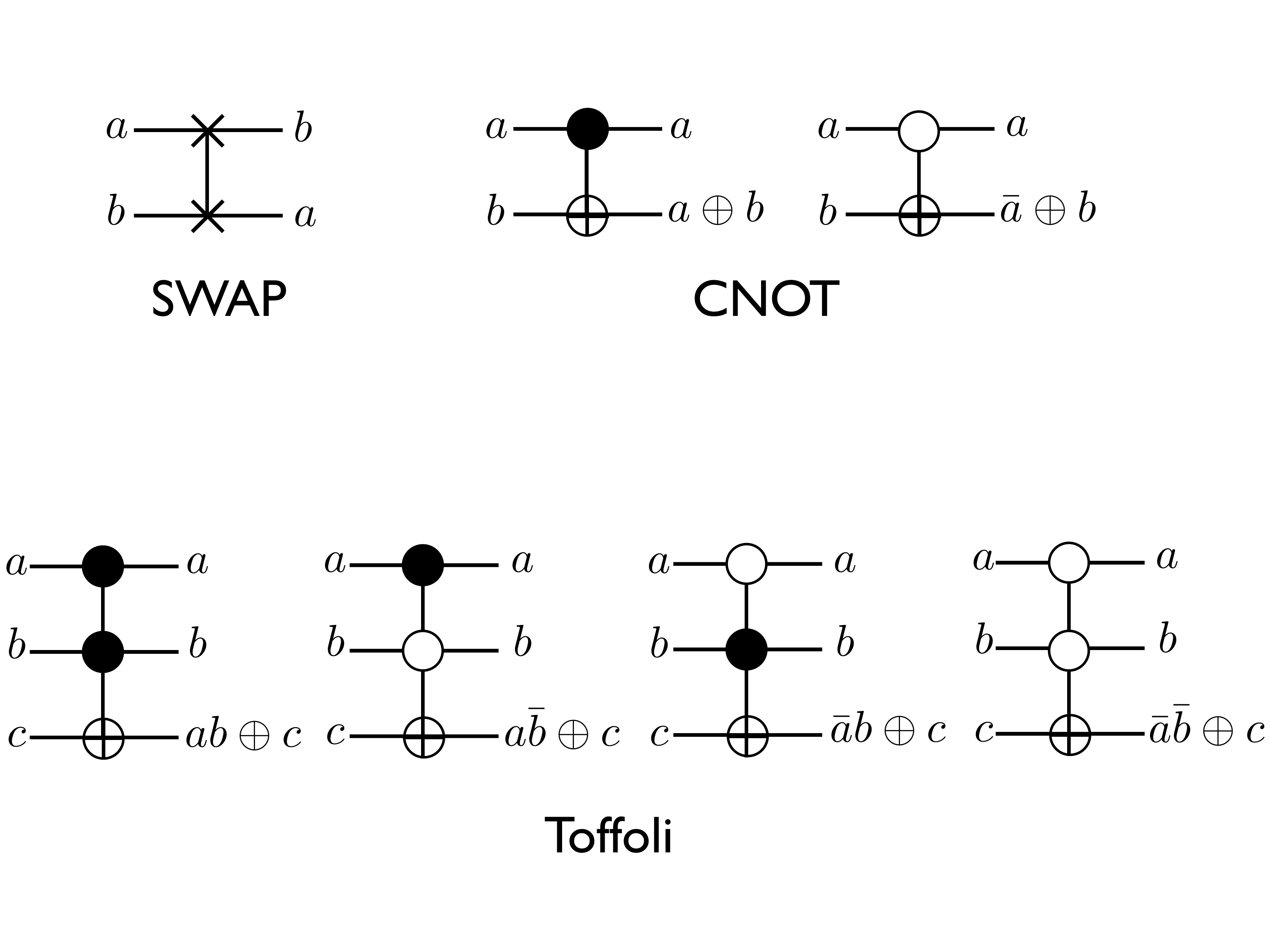}}
\caption{Reversible gates used in the quantum stochastic
  evolutions. The output values of the gates are defined when $a$,
  $b$, and $c$ take values 0 and 1. Top row: Two-qubit gates SWAP and
  CNOT. Bottom row: The different variations of the three-qubit
  Toffoli gates. We note that different variations of the CNOT and
  Toffoli gates can be obtained from one fixed variation plus NOT
  gates.}
\label{fig:gates}
\end{figure}

The quantum system we consider contains $n$ qubits and evolves
unitarily from an initial factorized state of the form $|\Psi_0\rangle
= |\psi_1\rangle \otimes |\psi_2\rangle \otimes \cdots \otimes
|\psi_n\rangle$ where each single-qubit state is defined as
$|\psi_j\rangle=\cos({\theta_j}/{2}) |0\rangle + \sin ({\theta_j}/{2})
e^{i\phi_j}|1\rangle$, with $\theta_j$ and $\phi_j$
arbitrary. Formally, the evolution is obtained by applying a unitary
matrix $U$ to the state vector, $|\Psi_{t}\rangle = U
|\Psi_{0}\rangle= \sum_{x} \Psi_t(x)\, |x\rangle$, where the states
$|x\rangle\equiv|x_1\,x_2\dots x_n\rangle$ form the computational
basis, with $x_j=0,1$ for $j=1,\dots,n$. Using the language of quantum
computing, we assume that this unitary matrix is represented by
gates. We recall that the two-qubit CNOT gate and arbitrary one-qubit
rotations are sufficient for universal quantum computing
\cite{divincenzo95}. In what follows, we shall restrict the gates to
the permutation group, which is a subgroup of the unitary group. The
restriction to the permutation subgroup of unitary transformations
allows for a much more efficient computation of the state of the
system as it evolves with gates. In particular, we consider the
unitary gates in the set $\mathcal I_3 =\{\mbox{SWAP}, \mbox{CNOT},
\mbox{Toffoli}\}$ depicted in Fig.~\ref{fig:gates}. We build a
stochastic quantum circuit $U=\prod_k^M U_k$ by drawing randomly with
uniform probability pairs or triplets of qubits and a random gate
$U_k\in\mathcal I_3$ with probability $1/3$. We remark that the
Toffoli gate alone is sufficient for universal classical computation
\cite{toffoli}. We also consider more restricted (and nonuniversal)
circuits obtained by employing only gates in the set $\mathcal I_2=\{
\mbox{SWAP}, \mbox{CNOT}\}$.

At each step $k$ of the circuit, the $n$ qubits are partitioned into
subsystems $(A,B)$ with $n_A$ and $n_B$ qubits, and the entanglement
properties of the system are obtained through the singular values
$\lambda_k>0$, $k=1\ldots,r$, which result from the Schmidt
decomposition \cite{ekert,peres} of the state $|\Psi_{t}\rangle =
\sum_{k=1}^r \lambda_k\; |\psi^{A}_{t\,(k)}\rangle \otimes
|\psi^{B}_{t\,(k)}\rangle$. The reduced density matrices $\rho_A=
\mbox{tr}_B (|\Psi_{t}\rangle \langle\Psi_{t}|)$ and $\rho_B=
\mbox{tr}_A (|\Psi_{t}\rangle \langle\Psi_{t}|)$ have eigenvalues
$\{p_k=\lambda_k^2\}$. These $p_k$ define a probability distribution
whose R\'enyi entropies are defined as \cite{renyi61}
\begin{equation}
\label{eq:S-lambda}
S_q (n_A,n_B)= \frac{1}{1-q}\, \log_2 \sum_{k=1}^r p_k^{q}\, ,
\end{equation}
with $\sum_{k=1}^r p_k=1$. The zeroth R\'enyi entropy is related to
the rank, namely, the number $r$ of nonzero singular values, $S_0 =
\log_2 r$. The $q=1$ R\'enyi entropy is the Shannon entropy measuring
the amount of information in the distribution $\{p_k\}$: $S_1=
-\sum_{k} p_k \log_2 p_k$.

\begin{widetext}

\begin{figure}[ht]
\centering
\scalebox{.48}{\includegraphics{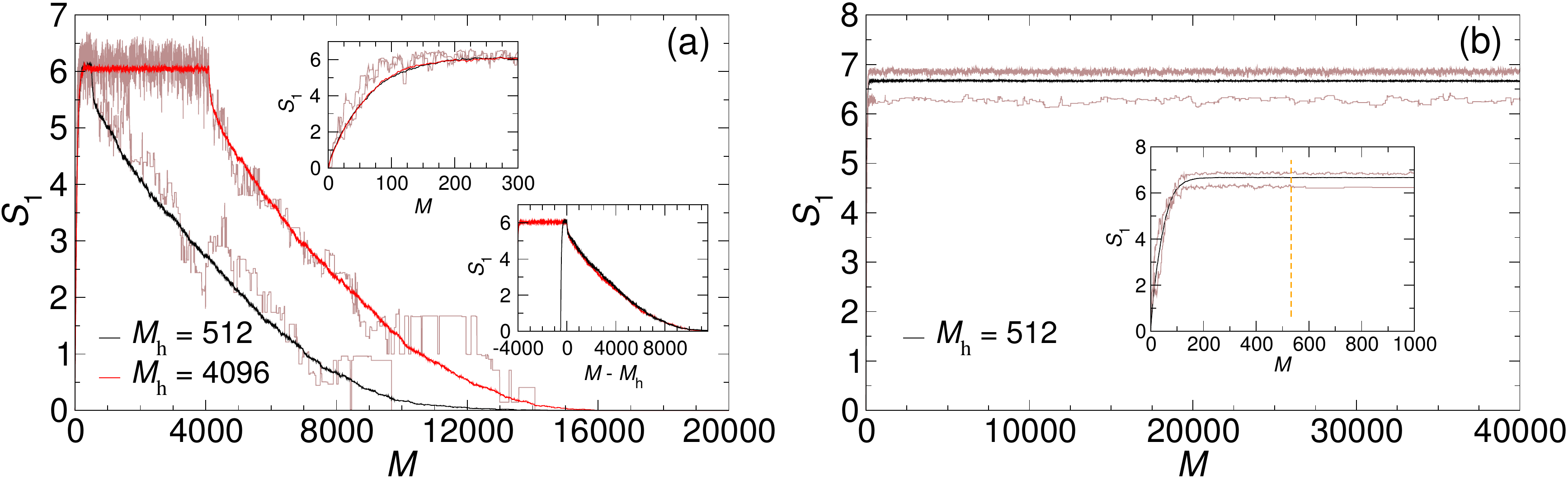}}
\caption{(Color online) Evolution of the entanglement entropy $S_1$
  obtained from the bipartition of the qubit string at the middle
  ($n_A=n_B=8$) as function of the number of applied gates. First,
  $M_{\rm h}$ reversible gates are randomly applied (heating period);
  second, a Metropolis algorithm is used to reverse the evolution and
  restore zero entropy (cooldown period). The solid black ($M_{\rm
    h}=512$) and red ($M_{\rm h}=4096$) lines result from averaging
  $S_1$ over 128 initial-state realizations. The brown lines show the
  evolution of $S_1$ for two typical realizations.
  (a) Only two-qubit gates are used. Upper inset: The heating
  period. Lower inset: Shifted curves, showing that the cooling, on
  average, is independent of the duration of the heating period. (b) A
  mixture of two-qubit and Toffoli gates is used. Inset: Detail of the
  transition between heating and cooling periods. The dashed line
  indicates the transition point.}
%
\label{fig:heat-cool}
\end{figure}

\end{widetext}

What happens to entanglement during the evolution with gates? One can
show that, under a generic stochastic random circuit, entanglement
grows linearly with time, and then saturates to its maximum possible
value \cite{asz,chamon2012}. This occurs typically, meaning that the
probability of having a different outcome is zero in the thermodynamic
limit. A similar behavior is obtained also for the restricted quantum
evolutions considered here, whether one uses two- or three-qubit
gates. The saturation value is typically reached after about $M\sim
n^2$ transformations. In Fig. \ref{fig:heat-cool}, we see a numerical
simulation of the protocol used, with both two-qubit and three-qubit
gates, which confirms this scenario. We call this part of the protocol
``entanglement heating.''

Because entanglement increases with the number of gates, in order to
revert the evolution to return back to the initial state, it is
natural to attempt an algorithm that completely disentangles the
system. The entanglement entropies provide a natural metric to use in
a minimization process. If one is able to remove all the entanglement
while recording the moves that led to the decreases, one builds one
possible reverse algorithm that takes the system from the final state
back to the initial (product) state. In practice, we implement such
disentangling or ``entropy cooling'' algorithm as follows. We attempt
a gate, chosen at random, and compute the change in entanglement
entropy. Then we decide whether or not to accept this gate into the
sequence according to a Metropolis algorithm: if the entanglement goes
down, we always take this move; if not, we take it with a certain
probability, which we decrease as function of the number of attempts
(similarly to simulated annealing, but applied to entanglement entropy
and not energy). More precisely, we use as the optimization function
the sum of the entanglement entropies over all bipartitions of the
system into $n_A$ and $n_B$ consecutive qubits with $n_A+n_B=n$,
namely, ${\cal S}_q = \sum_{n_A=1}^{n-1} S_q(n_A,n_B)$. The reason for
this choice is that a single bipartition is sensitive only to gates
that act on qubits in both subsystems $A$ and $B$. However, if one
considers the sums over the entanglement for all bipartitions, one is
sensitive to all reductions in entanglement, no matter where the gates
act.

The resulting $q=1$ R\'enyi entropy as a function of the gate number
for a given sequence of the algorithm using {\em only} the gates in
$\mathcal I_2$ is shown in Fig.~\ref{fig:heat-cool}(a). The data show
two examples of the entanglement evolution for two particular
``heating'' and ``cooling'' runs, as well as the average of 128
different realizations with random initial product states for 16
qubits, with each $\theta_j$ randomly picked from the interval
$[0,\pi]$, and $\phi_j=0,\pi$ (thus focusing on real wave
functions). We show data for the case when the system is entangled
with 512 and with 4096 gates. The system is ``cooled'' by minimizing
${\cal S}_0$ (similar results are obtained when minimizing ${\cal
  S}_2$). Notice that the ``cooling'' time for the average curve does
not depend on how long the system was ``heated,'' provided that the
same maximum entanglement entropy is reached. The disentangling
algorithm works for {\it all} individual realizations of the
protocol. We were {\it always} able to reverse to a completely
factorized tensor product state with zero entanglement. This is quite
remarkable, because the success of the algorithm does not depend at
all on the amount of the entanglement produced. So one may wonder
whether {\em every} quantum circuit can be reversed with such a
cooling protocol.

\begin{figure}
\centering \scalebox{.3}{\includegraphics{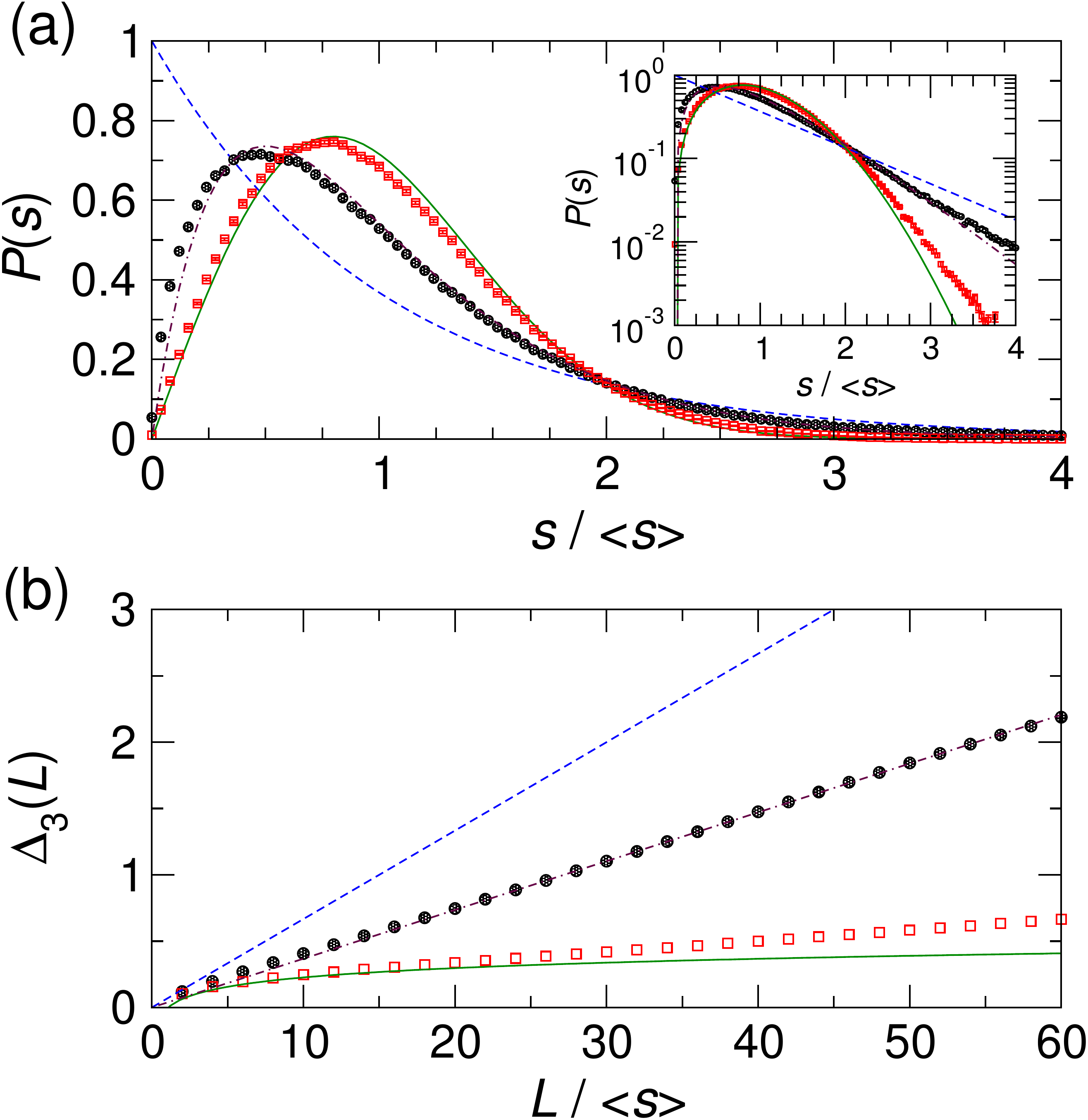}}
\caption{(Color online) (a) The distribution of the spacing between
  consecutive unfolded singular values obtained from a bipartition at
  the middle of a $n=16$ qubit string at the end of the heating period
  ($M_{\rm h} =512$). Solid black circles are for two-qubit gate
  heating and open red squares are for heating with a mixture of
  two-qubit and Toffoli gates.
  Solid green line: GOE prediction. Dashed blue line: Poisson
  distribution. Dotted-dashed maroon line: Semi-Poisson
  distribution. Inset: The tail of the distributions. (b) The average
  spectral rigidity $\Delta_3(L)$ obtained from the same spectra (a
  linear fit is used for the semi-Poisson line). A total of 5000
  realizations were used to compute the averages.}
\label{fig:stat_cont}
\end{figure}

To answer this question, consider now the case when entanglement
entropy ``heating'' involves the gates in $\mathcal I_3$. Then, apply
the disentangling algorithm using the same set of gates. For {\it all}
realizations studied, we find that it is {\it never} possible to
completely disentangle the state using the Metropolis protocol
described above \cite{note1}. In Fig. \ref{fig:heat-cool}(b) we show
two typical realizations of the heating and cooling protocol, with a
random initial product state and 512 random gate sequences for the
heating phase. We also show the average over 128 realizations.

And yet, by only looking at the amount of entanglement generated upon
``heating,'' we cannot tell whether the evolution is reversible by the
cooling algorithm. As we have shown, by heating with either two-qubit
gates or a mixture of two-qubit and Toffoli gates, one rapidly reaches
an almost maximally entangled state. Nevertheless, only for the former
are we able to reverse the system back into a product-state
form. Indeed, it is known that most states in the Hilbert space are
maximally entangled, and that generic quantum evolutions will
eventually lead to an almost maximally entangled state \cite{page,
  emerson, viola, asz}. This happens even under quantum quench with an
integrable Hamiltonian \cite{calabrese}.

What is in the entanglement, which is not the entanglement entropy,
that tells us whether a quantum evolution is reversible or not? The
answer lies in the statistics of the levels $\{p_k\}$ in the
entanglement spectrum. We have computed the entanglement spectrum of
the qubit string at the end of the heating period. The spectrum is
obtained from the singular values resulting from the Schmidt
decomposition of the quantum state upon bipartitioning of the qubit
string in the middle (i.e., $n_A=n_B=n/2)$. The spectrum is first
unfolded to yield a constant density before the statistical analysis
is performed (see the Appendix for a detailed description of the
unfolding procedure). In Fig. \ref{fig:stat_cont}(a) we show the
distribution of the spacings between adjacent singular values for
reversible cases (heating period performed with $\mathcal I_2$ gates)
and irreversible ones (heating period performed with $\mathcal I_3$
gates). The difference is striking: while the data points for the
irreversible case match quite closely the distribution of spacings of
the Gaussian orthogonal ensemble (GOE) of random matrices
\cite{mehtabook}, the data points for the reversible case show a
weaker repulsion and follow the so-called semi-Poisson statistics,
which has been proposed for the energy spectra of systems at
metal-insulator transitions \cite{bogomolny}. The difference in
behavior is also manifest in the spectral rigidity function
$\Delta_3(L)$, which measures, for a given interval $L$, the
least-square deviation of the spectral staircase from the best-fitting
straight line \cite{dyson1963}. In Fig. \ref{fig:stat_cont}b,
long-range correlations are much stronger in the irreversible case,
with the data points also falling close to the GOE prediction. For the
reversible case, the singular values are much less correlated and the
spectrum much less rigid. This indicates that the statistical
fluctuations of the entanglement spectra of irreversible systems are
similar to those observed in the energy spectrum of the so-called
quantum chaotic systems \cite{gutzwiller}.

\begin{figure}
\centering \scalebox{.3}{\includegraphics{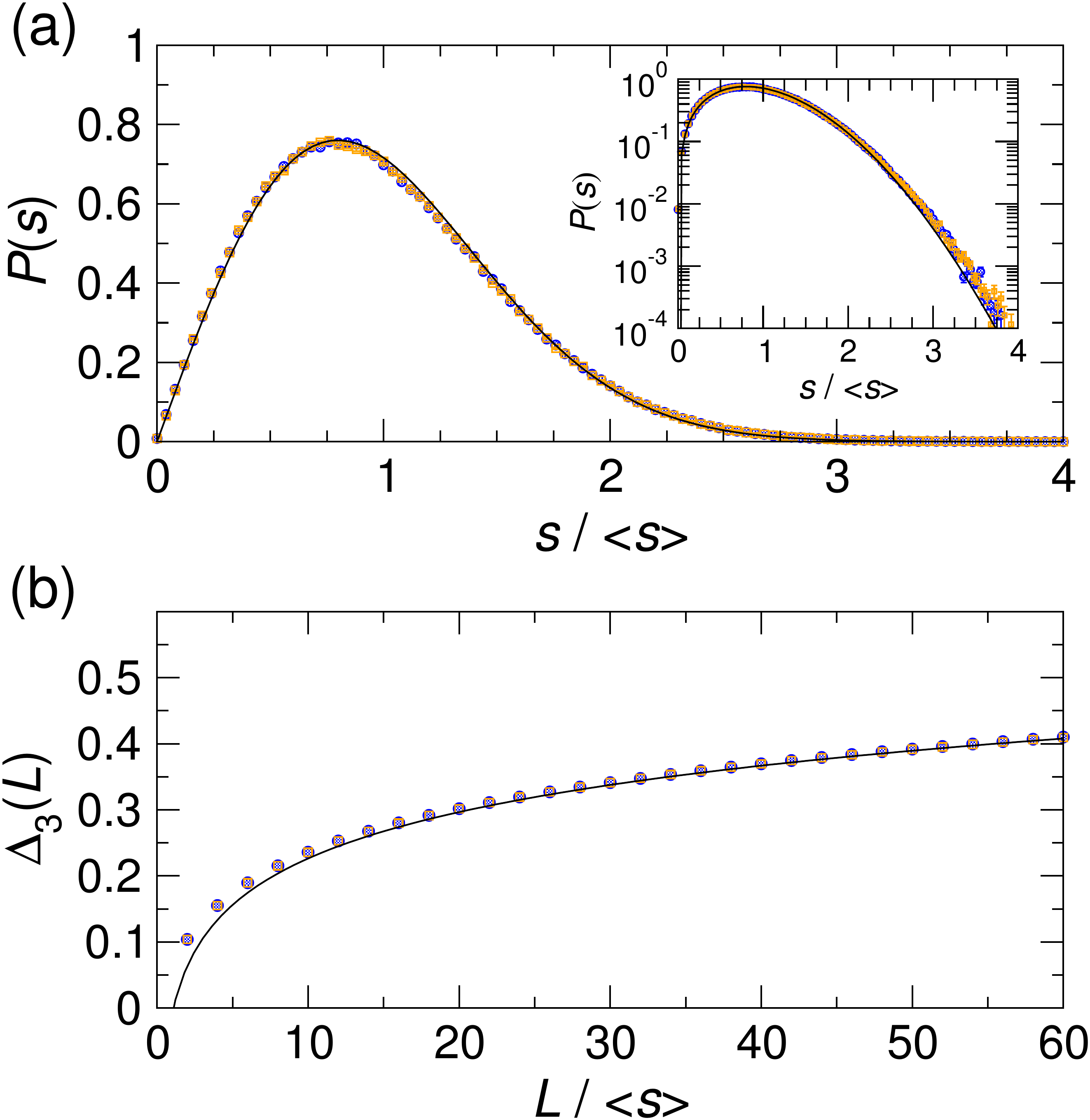}}
\caption{(Color online) The statistical fluctuations of the
  entanglement spectrum for initial product states
  $|\chi^{(1)}\rangle$ (orange empty squares) and $|\chi^{(2)}\rangle$
  (blue full circles) entangled with a mixture of two-qubit and
  Toffoli gates ($n=16$, $M_{\rm h} = 512$). The solid green line is
  the GOE prediction. (a) Distribution of unfolded singular value
  spacings. Inset: distribution tails. (b) Average spectral rigidity
  $\Delta_3(L)$. Statistical averages performed over 5000
  realizations.}
\label{fig:stat_disc}
\end{figure}

In Fig. \ref{fig:stat_disc} we also show the entanglement level
statistics for the particular case when one starts with initial
factorized states of the form $|\chi^{(k)}\rangle
=|\psi_1^{(k)}\rangle \otimes |\psi_2^{(k)}\rangle \otimes \cdots
\otimes |\psi_n^{(k)}\rangle$, $k=1,2$, where $|\psi_1^{(1)}\rangle =
|0\rangle$ and $|\psi_j^{(1)}\rangle = (|0\rangle+|1\rangle)/\sqrt{2}$
for $j=2,\ldots,n$, and $|\psi_j^{(2)} \rangle =
(|0\rangle-|1\rangle)/\sqrt{2}$ for $j=1,\ldots,n$. We evolve these
$n=16$-bit states with $M=512$ gates chosen randomly from the set
$\mathcal I_3$. The data in Fig. \ref{fig:stat_disc} clearly conform
to the GOE statistics, and we observe that the disentangling algorithm
again fails, indicating that reversing the computation is extremely
difficult.

In quantum mechanics, irreversibility, chaos, nonintegrability and
thermalization are phenomena often associated with one
another. Unfortunately, some of these notions are ill defined, such as
integrability and lack thereof, and the associations are either weak
or plagued by counterexamples. For instance, irreversibility can be
associated to both chaotic and nonchaotic Hamiltonians \cite{znidaric,
  benenti} and there are nonintegrable systems that do not thermalize
\cite{eisert}. Moreover, some of these concepts are only defined in
the context of time-independent Hamiltonian evolutions. For instance,
the energy levels of a chaotic Hamiltonian show Wigner-Dyson
statistics.

In this Letter we presented an alternative approach to the question of
irreversibility and complex behavior in quantum systems that works
purely at the wave-function level. We did so by studying the
eigenvalues of the reduced density matrix of a subsystem, the
so-called entanglement spectrum \cite{footnote}. We showed that (i) a
disentangling Metropolis algorithm provides a firm notion of
reversibility, namely, the evolution can be inverted if the state can
be disentangled, and (ii) irreversibility arises when the level
statistics of the entanglement spectrum of a subsystem is
Wigner-Dyson.

On the other hand, in the example we studied where the spectrum did
not follow Wigner-Dyson statistics, we were always capable of
reverting the evolution, even with zero knowledge about the quantum
circuit. It is remarkable that the length of the reverted circuit does
not depend on the length of the initial circuit, as long as the
maximum entanglement entropy is reached. In the disentangling
algorithm, we obtained similar results with the R\'enyi entropy
$S_2$. This is remarkable because $S_2$ is an observable that can be
measured \cite{s1,s2b}, for example, in optical lattices with
ultracold atomic gases \cite{bloch}.

The results of this work motivate several questions and
applications. The method presented here is applicable to any kind of
quantum evolution, regardless of whether it comes from a quantum
circuit, a time-dependent or -independent Hamiltonian system, or an
open quantum system. First of all, we can examine the behavior of the
entanglement level spacing statistics in integrable Hamiltonian models
both in the ground state or during the time evolution after a quantum
quench. Using techniques such as the density matrix renormalization
group, one can study these models once integrability is broken. We
believe that our approach can shed new light on the notion of
integrability and lack thereof in quantum systems. The possibility of
studying quantum systems away from equilibrium and their universal
properties in dynamical phase transitions \cite{polkovnikov} and
many-body localization \cite{boris, huse, refael} is another feature
of the method that only involves wave functions. Similarly, we can
study the behavior of the entanglement level spacing statistics at
critical points of integrable and nonintegrable systems
\cite{entmanybody, arul}. The adiabaticity of time-dependent quantum
processes \cite{zurek} can be examined under the lens of the
entanglement spectrum as well, with potential applications to
adiabatic quantum computing. Moreover, one can study how the
complexity of the entanglement spectrum is related to quantum
algorithms capable of giving an exponential speedup
\cite{arul2}. Under the same lens of the entanglement spectrum, one
should study the typicality of quantum chaos in random states
\cite{vinayak, zic}. Entanglement is very ubiquitous in the Hilbert
space \cite{zic, emerson}, and while this feature has been crucial to
show the typicality of thermalization in closed quantum systems
\cite{nature}, this also means that entanglement entropy is unable to
characterize quantum irreversibility, and the difference between
integrable and nonintegrable systems. Our results show that the
understanding of complex quantum behavior lies in the statistics of
the fluctuations of the entanglement level spacing.

We acknowledge financial support from the U.S. National Science
Foundation through grants CCF 1116590 and CCF 1117241, and by the
National Basic Research Program of China Grant 2011CBA00300,
2011CBA00301, the National Natural Science Foundation of China Grant
61033001, 61361136003.



\newpage

\onecolumngrid


\section*{Appendix: Spectral analysis}

Before computing the statistical properties of the singular values
${\lambda_k}$, ${k=1,\ldots,d}$, the spectrum must be unfolded in such
a way to produce a new sequence ${s_k=f(\lambda_k)}$,
${k=1,\ldots,d}$, with a uniform density. In Fig. 1 we show typical
spectra obtained for $n=16$ qubit systems after a heating period of
512 gates and starting from a random initial state. For the cases
where the initial state has amplitudes $W(x)$ taking continuous values
(in the set of real numbers), both small and large jumps appear in
some spectra, particularly for the case of circuits involving only
two-bit permutation gates. We thus divided the spectrum of each
realization into segments and fitted to each segment an independent
polynomial function (see Fig. 1). Segments shorter than 10 singular
values were not considered and singular values near the beginning or
the end of the spectrum were discarded. For the cases of initial
states with discrete amplitudes, namely, $W(x)=0,1$ or $W(x)=\pm 1$,
the spectra are quite smooth (see Fig. 2) and follow accurately a
Marchenko-Pastur distribution, which can be derived from the
semi-circle eigenvalue distribution of Random Matrix Theory:
\begin{eqnarray}
s_k & = & \frac{4}{\pi} \int_{x_k}^1 dx \, \sqrt{1-x^2} \\ & =
& 1 - \frac{2}{\pi} \left[ x_k\sqrt{1-x_k^2} + {\rm arcsin}(x_k)
  \right],
\end{eqnarray}
where $x_k^2 = \lambda_k^2\,d/4Z(1-p_1)$ and $Z=\sum_{k=1}^d
\lambda_k^2$. Here, $p_1$ is the probability of choosing an initial
amplitude $W=1$. In these cases, the unfolding was done with the same
continuous curve for all realizations.

The distribution of spacings, $P(s)$, where $s=s_{k+1}-s_k$, accounts
for short-range correlation and repulsion. Results were compared to
the Poisson and GOE predictions (we did not need to consider other
ensembles because only the entanglement entropy of states with real
amplitudes were analyzed): $P_{\rm Poisson}(s) = (1/\Delta)
\exp(-s/\Delta)$ and $P_{\rm GOE}(s) = \frac{\pi}{2}\, (s/\Delta^2)\,
\exp(-\pi s^2/4\Delta^2)$, where $\Delta = \langle s\rangle$.

The spectral rigidity was quantified through the function
\begin{equation}
\Delta_3(L) = \frac{1}{L} \left\langle {\rm min}_{a,b}
\int_{-L/2}^{L/2} dS\, \left[ N(S+S_0) - (a+bS)\right]^2
\right\rangle_{S_0},
\end{equation}
where
\begin{equation}
N(S) = \sum_{k=1}^d \theta(S-s_k)
\end{equation}
is the spectral staircase. Notice that for a given value of the
interval $L$, the averaging also involves sweeping over the spectrum
by varying the center point $S_0$. The results were compared with the
Poisson and GOE predictions, namely, $\Delta_3^{\rm Poisson}(L) =
L/(15\Delta)$ and $\Delta_3^{\rm GOE}(L) = \ln(L/\Delta)/\pi^2 -
0.00696$ for $L\gg\Delta$.

\begin{figure}
\centering
\scalebox{.5}{\includegraphics{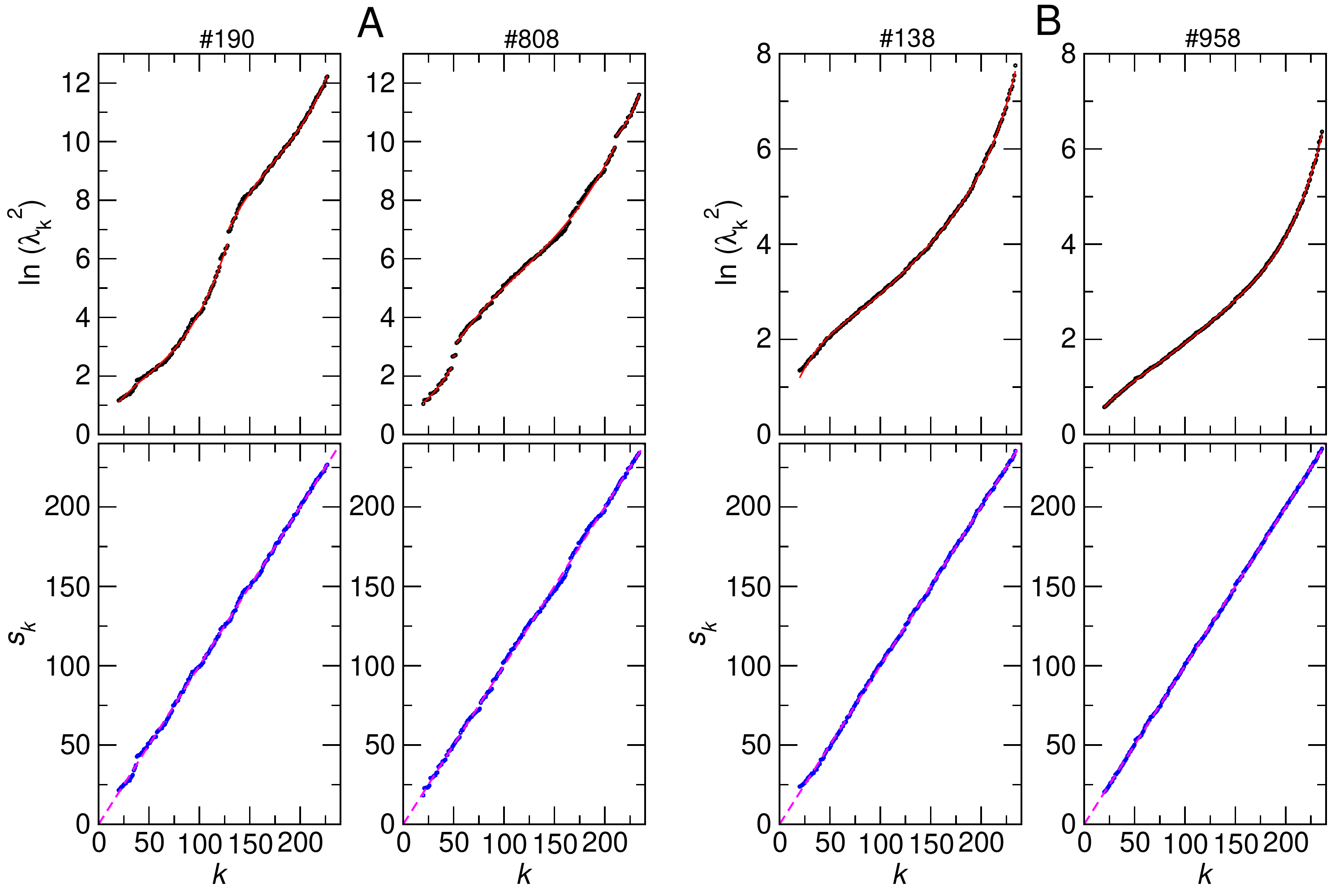}}
\caption{(a) Upper panels: examples of the entanglement spectra
  obtained by evolving a $n=16$ qubit system with 512 random two-bit
  reversible gates starting from two different initial states with a
  continuous distribution of amplitudes (sample numbers are indicated
  at the top of the graphs). Black circles represent the raw data and
  the solid red line is the result of a multi-segment polynomial fit
  (third degree). Lower panels: the blue circles are the resulting
  unfolded spectrum. The magenta dashed line is a guide to the eye,
  showing how close the distribution is to a straight line. (b)
  Similar to (a), but for states evolved with a mixture of reversible
  two- and three-bit (Toffoli) gates.}
\end{figure}



\begin{figure}
\centering
\scalebox{.5}{\includegraphics{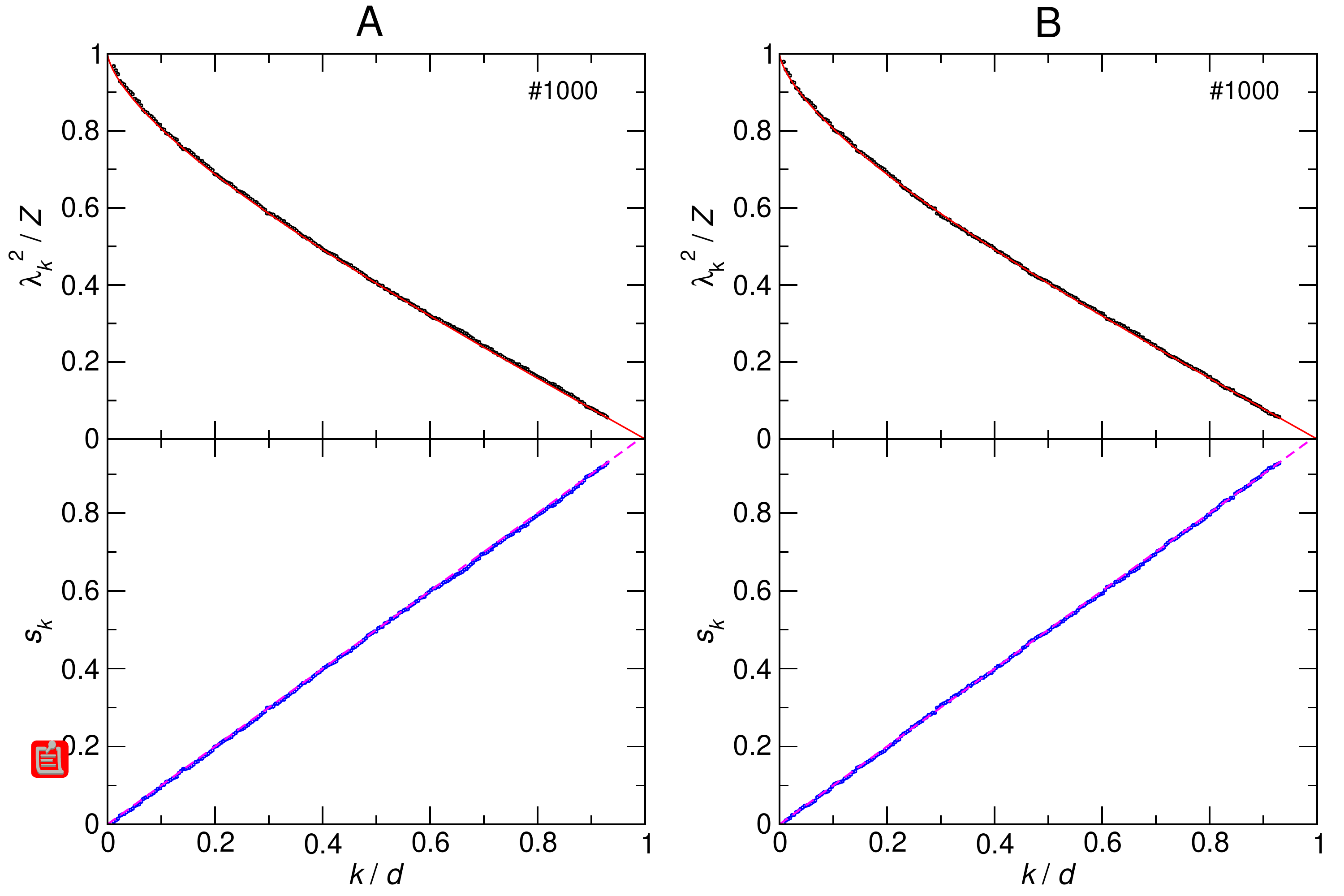}}
\caption{(a) Upper panels: example of the entanglement spectra
  obtained by evolving a $n=16$ ($d=256$) qubit system with a mixture
  of 512 random reversible two- and three-bit (Toffoli) gates starting
  from an initial states with a discrete distribution of amplitudes,
  namely, $W(x)=0,1$. Black circles represent the raw data and the
  solid red line indicates $k/d = 1-(2/\pi)[x_k\sqrt(1-x_k^2)-{\rm
      asin}(x_k)]$, with $x_k=\lambda_k^2/Z$ and $Z=\sum_{k=1}^d
  \lambda_k^2$. Lower panels: the blue circles are the resulting
  unfolded spectrum. The magenta dashed is a guide to the eye, showing
  how close the distribution is to a straight line. (b) Similar to
  (a), but for initial states with a distribution of amplitudes
  $W(x)=\pm 1$.}
\end{figure}



\begin{thebibliography}{99}

\bibitem{caux} J.-S. Caux and J. Mossel, J. Stat. Mech. (2011) P02023.

\bibitem{Peres84} A. Peres, 
Phys. Rev. A {\bf 30}, 1610 (1984).

\bibitem{gutzwiller} M. C. Gutzwiller, {\it Chaos in Classical and
  Quantum Mechanics} (Springer Verlag, New York, 1991).

\bibitem{znidaric} T. Gorin, T. Prosen, T.H. Seligman, and
  M. \v{Z}nidari\v{c}, Phys. Rep. {\bf 435}, 33 (2006).

\bibitem{benenti} G. Benenti and G. Casati, \pre {\bf 79}, 025201(R)
  (2009).

\bibitem{divincenzo95} D. P. DiVincenzo, Phys. Rev. A {\bf 51}, 1015
  (1995).

\bibitem{toffoli} E. Fredkin and T. Toffoli, Int. J.
  Theor. Phys. {\bf 21}, 219 (1982).

\bibitem{ekert} A. Ekert and P. L. Knight, 
Am. J. Phys. {\bf 63}, 415 (1995).

\bibitem{peres} A. Peres, {\it Quantum Theory: Concepts and Methods}
  (Kluwer Academic, Dordrecht, 1995).

\bibitem{renyi61} A. R\'enyi, in
  {\it Proceedings of the 4th Berkeley Symposium on Mathematical
    Statistics and Probability, Vol. 1} (University of California
  Press, Berkeley, 1961), p. 547.

\bibitem{asz} A. Hamma, S. Santra, and P. Zanardi 
Phys. Rev. Lett. {\bf 109}, 040502 (2012).

\bibitem{chamon2012} C. Chamon and E. R. Mucciolo, 
Phys. Rev. Lett. {\bf 109}, 030503 (2012).

\bibitem{note1} The gates used in $\mathcal I_3$ are just a subgroup
  of the full unitary group and are not universal. A circuit
  comprising a universal set of gates for quantum computation--such
  CNOT and general one-qubit rotations--produces states that also fail
  to be disentangled. A systematic study of the full unitary group
  will be presented elsewhere.

\bibitem{page} D. N. Page, Phys. Rev. Lett. {\bf 71}, 1291 (1993).

\bibitem{emerson} J. Emerson {\it et al.}, Science {\bf 302}, 2098
  (2003).

\bibitem{viola} W. G. Brown, Y. S. Weinstein, and L. Viola,
  Phys. Rev. A {\bf 77}, 040303(R) (2008).

\bibitem{calabrese} P. Calabrese and J. Cardy, J. Stat. Mech.  (2005),
  P04010.

\bibitem{bogomolny} E. B. Bogomolny, U. Gerland, and C. Schmit,
  Phys. Rev. E {\bf 59}, R1315 (1999).

\bibitem{mehtabook} M. L. Mehta, {\it Random Matrices}, 3rd. edition
  (Academic Press, Amsterdam, 2004).

\bibitem{dyson1963} F. J. Dyson and M. L. Mehta, 
J. Math. Phys. {\bf 4}, 701 (1963).

\bibitem{eisert} C. Gogolin, M. P. M\"uller, and J. Eisert,
  Phys. Rev. Lett. {\bf 106}, 040401 (2011).

\bibitem{footnote} {Many authors use this terminology for the
  logarithms of such eigenvalues, particularly in the context of
  entangling Hamiltonians in condensed matter systems.}

 \bibitem{s1} P. Horodecki and A. Ekert, 
Phys. Rev. Lett. {\bf 89}, 127902 (2002).

\bibitem{s2b} D. A. Abanin and E. Demler, 
Phys. Rev. Lett. {\bf 109}, 020504 (2012).

\bibitem{bloch} M. Greiner, O. Mandel, T. Esslinger, T. W. Hansch, and
  I. Bloch,
Nature (London) {\bf 415}, 39 (2002).

\bibitem{polkovnikov} A. Polkovnikov, A. K. Sengupta, A. Silva, and
  M. Vengalattore, 
Rev. Mod. Phys. {\bf 83}, 863 (2011).

\bibitem{boris} D. M. Basko, I. L. Aleiner, and B. L. Altshuler, 
Ann. Phys. {\bf 321}, 1126 (2006).

\bibitem{huse} A. Pal and D. A. Huse, 
Phys. Rev. B {\bf 82}, 174411 (2010).

\bibitem{refael} S. Iyer, V. Oganesyan, G. Refael, and D. A. Huse, 
Phys. Rev. B {\bf 87}, 134202 (2013).

\bibitem{arul} A. Lakshminarayan and V. Subrahmanyam, Phys. Rev. A
  {\bf 71}, 062334 (2005).

\bibitem{entmanybody} L. Amico, R. Fazio, A. Osterloh, and V. Vedral, 
Rev. Mod. Phys. {\bf 80}, 517 (2008).

\bibitem{zurek} H. T. Quan and W. H. Zurek,
New J. Phys. {\bf 12}, 093025 (2010).

\bibitem{arul2} K. Maity and A. Lakshminarayan, Phys. Rev. E {\bf 74},
  035203(R) (2006).

\bibitem{zic} S. J. Szarek, E. Werner, and K. Zyczkowski, J. Phys. A
  {\bf 44}, 045303 (2011).

\bibitem{vinayak} Vinayak and M. Znidaric, J. Phys. A {\bf 45}, 125204
  (2012).

\bibitem{nature} S. Popescu, A. J. Short, and A. Winter, 
Nature Phys. {\bf 2}, 754 (2006).

\end{thebibliography}
\end{document}